%% file: main.tex
\documentclass{iscram}

\usepackage{hyperref}
\hypersetup{colorlinks,allcolors=black}

\usepackage[utf8]{inputenc}
\bibliography{main}
\usepackage{lipsum}
\usepackage{tikz}
\usepackage{fancyvrb}
\usepackage{listings}
\usepackage{enumitem}
\usepackage{tcolorbox}
\usepackage{multicol}
\usepackage{siunitx}
\usepackage{multirow}
\usepackage{graphicx}
\usepackage{subfig}
\usepackage{colortbl}
\usepackage{tabto}

\usepackage{array}
\usepackage{booktabs, caption}
\usepackage{multirow, makecell}

\usepackage{lipsum}

\lstdefinestyle{common}{
  xleftmargin=.5em,
  xrightmargin=.5em,
  frame=single,framesep=.5em,framerule=0pt,
  fancyvrb=true,
  basicstyle=\ttfamily,
  keywordstyle=\color{cyan!50!blue!75!black}\bfseries,
  commentstyle=\color{red!50!black}\itshape,
  stringstyle=\ttfamily\color{green!50!black},
  numbers=none,
  showspaces=false,
  showstringspaces=false,
  fontadjust=true,
  keepspaces=true,
  flexiblecolumns=true,
  emphstyle=\color{red},
}
\lstdefinestyle{TeX}{
  style=common,
  backgroundcolor=\color{blue!5},
  aboveskip=5pt,
  belowskip=5pt,
  language=[LaTeX]TeX,
  moretexcs={
    abstract, addbibresource, iscramset, keywords, mainmatter,
    maketitle, printbibliography, subsection, subsubsection, url,
    urldef, href, includegraphics, ldots, parencite, citeauthor,
    citeyear, citetitle, midrule, toprule, bottomrule
  },
  fancyvrb=true,
}
\lstdefinestyle{console}{
  style=common,
  backgroundcolor=\color{gray!10},
  aboveskip=5pt,
  belowskip=5pt,
}

\newlist{options}{description}{1}
\setlist[options]{%
  beginpenalty=10000,%
  itemsep=.5\parskip plus .3\parskip minus .2\parskip,
  parsep=.5\parskip plus .3\parskip minus .2\parskip,
  topsep=.5\parskip plus .3\parskip minus .2\parskip,
  partopsep=.5\parskip plus .3\parskip minus .2\parskip,
  style=nextline,labelindent=1em,%
  font=\normalfont\ttfamily}

\colorlet{macro color}{cyan!50!blue!75!black}
\colorlet{option color}{red!50!black}
\colorlet{generic color}{green!40!black}

\newtcolorbox{pseudoTeX}{colback=blue!5,colframe=blue!5,before=\nobreak}
\let\LaTeXorig\LaTeX
\renewcommand\LaTeX{\bgroup\fontfamily{lmr}\selectfont\upshape\LaTeXorig\egroup}

\urldef{\sitewebgind}\url{http://gind.mines-albi.fr}
\urldef{\sitewebminesalbi}\url{http://www.mines-albi.fr}
\urldef{\gmailpaulgaborit}\url{paul.gaborit@gmail.com}
\urldef{\jdmail}\url{j.doe@example.com}
\urldef{\sitex}\url{www.example.com}

\iscramset{
  CoRe Paper 2020={Social Media for Disaster Response and Resilience},
 title={Rapid Damage Assessment Using Social Media Images by Combining Human and Machine Intelligence
   },
  short title={Rapid Damage Assessment Using Social Media Images},
  author={
    short name={Imran},
    full name=Muhammad Imran\thanks{Corresponding author},
    affiliation={
      Qatar Computing Research Institute\\
      Hamad Bin Khalifa University\\
      Doha, Qatar\\
      mimran@hbku.edu.qa
    },
  },
  author={
    full name=Firoj Alam,
    affiliation={
      Qatar Computing Research Institute\\
      Hamad Bin Khalifa University\\
      Doha, Qatar\\
      fialam@hbku.edu.qa 
    },
  },
  author={
    full name=Umair Qazi,
    affiliation={
      Qatar Computing Research Institute\\
      Hamad Bin Khalifa University\\
      Doha, Qatar\\
      uqazi@hbku.edu.qa 
    },
  },
  author={
    full name=Steve Peterson,
    affiliation={
      Montgomery County, Maryland Community Emergency Response Team\\	
      United States\\
      stevepeterson2@gmail.com 
    },
  },
  author={
    full name=Ferda Ofli,
    affiliation={
      Qatar Computing Research Institute\\
      Hamad Bin Khalifa University\\
      Doha, Qatar\\
      fofli@hbku.edu.qa 
    },
  },
}

\usepackage{tikzpagenodes}
\usetikzlibrary{fit}

\begin{document}

\maketitle

\makeatletter
\makeatother

\abstract{Rapid damage assessment is one of the core tasks that response organizations perform at the onset of a disaster to understand the scale of damage to infrastructures such as roads, bridges, and buildings. This work analyzes the usefulness of social media imagery content to perform rapid damage assessment during a real-world disaster. An automatic image processing system, which was activated in collaboration with a volunteer response organization, processed ${\sim}$280K images to understand the extent of damage caused by the disaster. The system achieved an accuracy of 76\% computed based on the feedback received from the domain experts who analyzed ${\sim}$29K system-processed images during the disaster. An extensive error analysis reveals several insights and challenges faced by the system, which are vital for the research community to advance this line of research.}

\keywords{social media, damage assessment, artificial intelligence, image processing}

\input{introduction}

\input{related_work}

\input{deployment_details}
\input{results}

\input{discussion_conclusion}

\printbibliography

\end{document}

%% file: introduction.tex
\section{Introduction}
\label{sec:introduction}

Rapid damage assessment is a task that humanitarian organizations perform within the first 48 to 72 hours of a disaster and is considered a prerequisite of many disaster management operations\footnote{\url{https://www.fema.gov/media-library/assets/documents/109040}}. Assessing the severity of damage helps first responders understand affected areas and the extent of impact for the purpose of immediate rescue and relief operations. Moreover, based on the results of early damage assessment, humanitarian organizations identify focus areas to make detailed assessment for long-term relief and rehabilitation of the affected population\footnote{\url{http://www.resiliencenw.org/2012files/LongTermRecovery/DisasterAssessmentWorkshop.pdf}}. However, traditional ways to perform rapid damage assessment require sending experts to the disaster affected area to conduct field assessments that include taking pictures of the damaged infrastructure, interviewing people, and collecting relevant data from other reliable sources. These experts perform analysis and interpretation of the gathered data before writing a report for planners and decision-makers. Limited human resources and severe living conditions in the disaster area are only a few examples of challenges facing field assessment experts. Such challenges can delay data gathering, damage assessment, and ultimately, relief operations.

Past works on rapid damage assessment used Synthetic Aperture Radar (SAR), remote sensing, and satellite image processing techniques~\parencite{plank2014rapid, barrington2012crowdsourcing, pesaresi2007rapid}. These approaches use costly data sources and are time consuming to deploy and collect relevant data. Furthermore, satellite data is susceptible to noise such as clouds, especially during weather-induced disasters, e.g., hurricanes. The focus of this work is to analyze the usefulness of non-traditional data sources such as social media to perform rapid damage assessment. More specifically, as opposed to textual content for damage detection~\parencite{kryvasheyeu2016rapid}, we are interested in the imagery content shared during an ongoing natural disaster to identify images that contain damage caused by the disaster. 

Microblogging and social media platforms such as Twitter play an increasingly important role during disasters~\parencite{castillo2016big,imran2015processing}. People turn to Twitter to get updates about an ongoing emergency event~\parencite{starbird2010chatter,hughes2009twitter}. More importantly, when people in disaster areas share information about what they witness in terms of damages caused by the disaster, flooded streets, reports of missing, trapped, injured or deceased people, or other urgent needs, that information could potentially be leveraged by humanitarian organizations to gain situational awareness and to plan relief operations~\parencite{imran2013extracting,purohit2014identifying}. 

In addition to the textual messages, images shared on Twitter carry important information pertinent to humanitarian response. This work focused on the real-time analysis of the imagery data shared on Twitter during Hurricane Dorian in 2019. In collaboration with a volunteer response organization, Montgomery County, Maryland Community Emergency Response Team (MCCERT)\footnote{We met the lead of the CERT team in one of the ISCRAM conferences and discussed the possibility to do a joint activation of our automatic image processing system for damage assessment.}, we activated our image processing system before Hurricane Dorian made landfall in the Bahamas. Based on the information requirements of our partner organization, the system filtered images that were relevant to the disaster and identified the ones that showed some damage content (e.g., damaged buildings, roads, bridges). More specifically, the damage analysis task assessed the severity of the damage using three levels: (i) severe damage, (ii) mild damage, and (iii) little-to-no damage (i.e., none).

During a 13-day deployment period, the system collected around ${\sim}$280K images. It used machine learning techniques to eliminate duplicate and irrelevant images before performing the damage assessment. As a result, around ${\sim}$160K images were found as relevant and around ${\sim}$26K as containing some damage content. Domain experts from our partner volunteer response organization examined an evolving sample of images during the disaster. The purpose of having human-in-the-loop was two-fold. \textit{First}, to keep an eye on the system generated output to verify the system was correctly classifying the images and make corrections if a mistake was identified. \textit{Second}, use the human corrections to better train the system for future deployments. 

The human experts performed two tasks while examining over ${\sim}$29K images over several days during the system's deployment period. \textit{First}, they determined if an image contained any damage content. \textit{Second}, if an image was identified as containing damage, they would determine the severity of the damage using the three severity levels mentioned above.

Based on the results of each expert's assessment, the system achieved an accuracy of 76\% for the damage detection task and 74\% for the damage severity assessment task. These are reasonable accuracy scores, which prove the effectiveness of the system for analyzing real-world disaster imagery data for rapid damage assessment. Furthermore, we performed an error analysis of the corrections resulting from performing the two tasks. Among common mistakes, we observed that the system is weak in identifying scenes that show flooding taken from afar, foggy or blurry scenes, and low-light scenes. Moreover, images that resembled damage scenes but were verified as incorrect confused the system. For example, a pile of trash would sometimes be confused as damage. Identifying deficiencies during our deployment not only helps us improve our machine learning models, but also provides valuable information for the crisis informatics research community to better understand challenges of analyzing social media imagery data during real-world disaster situations. This could lead to the discovery of additional methods and models that seek similar qualifying actionable machine output imagery to benefit decision-makers.

The rest of the paper is organized as follows. The next section summarizes \textit{Related Work}. In the \textit{Hurricane Dorian Deployment} section
we provide details of the event and our system deployment. Then, 
we report the data collection and analysis in the section \textit{Data and Results}. We later discuss our findings in the \textit{Discussion} section, identify challenges, and provide future directions.
Finally, we conclude the paper in the last section.








%% file: related_work.tex
\section{Related Work}
\label{sec:related_work}

The importance of imagery content for disaster response has been reported in a number of  studies~\parencite{TurkerM:IJRS04,chen2013understanding,plank2014rapid,FengT:NHESS14,fernandez2015uav,Nattari:DSAA17,erdelj2016uav,ofli2016combining}. These studies dominantly analyze aerial and satellite imagery data. 
For instance, \cite{TurkerM:IJRS04} analyze post-earthquake aerial images to detect damaged infrastructure caused by the August 1999 Izmit earthquake in Turkey. 
\cite{plank2014rapid} provides a comprehensive overview of multi-temporal Synthetic Aperture Radar procedures for damage assessment and highlights the advantages of SAR compared to the optical sensors. 

On the other hand, \cite{fernandez2015uav} and \cite{Nattari:DSAA17} report the importance of images captured by Unmanned Aerial Vehicles (UAV) for damage assessment while highlighting the limitations of remote sensing data. These studies propose per-building damage scores by analyzing multi-perspective, overlapping and high-resolution oblique images obtained from UAVs.
\cite{ofli2016combining} also highlights the importance of UAV images while addressing the limitations of satellite images. The authors propose a methodology that enables volunteers to annotate aerial images, which is then combined with machine learning classifiers to tag images with damage categories.

Very recently, the study of social media image analysis for disaster response has received attention from the research community~\parencite{daly2016mining,Mouzannar2018,alam2018SocialMedia}. 
For example, \cite{daly2016mining} analyze images extracted from social media data collected during a fire event. Specifically, they analyze spatio-temporal meta-data associated with the images and suggest that geo-tagged information is useful to locate the fire-affected areas. \cite{Mouzannar2018} investigate damage detection by focusing on human and environmental damages. Their study includes collecting multimodal social media posts and labeling them with six categories such as (1) infrastructural damage (e.g., damaged buildings, wrecked cars, and destroyed bridges) (2) damage to natural landscape (e.g., landslides, avalanches, and falling trees) (3) fires (e.g., wildfires and building fires) (4) floods (e.g., city, urban and rural) (5) human injuries and deaths, and (6) no damage. 

While many of the past works on rapid damage assessment need expensive data sources, some of which are also time consuming to deploy such as UAVs, satellites, and SAR, our work highlights the usefulness of Twitter images and utilizes an image processing pipeline proposed in~\parencite{nguyen2017automatic}. This image processing system filters irrelevant content, removes duplicates, and assesses damage severity for real-time damage assessment using deep learning techniques and human-in-the-loop.

%% file: deployment_details.tex
\section{Hurricane Dorian Deployment}
\label{sec:dorian_deployment}

\subsection{Hurricane Dorian}

On the morning of August 30, 2019, Hurricane Dorian was a Category 2 in the eastern Caribbean barreling toward the northern Bahaman Islands and central Florida. In the next 24 hours, the tropical storm rapidly intensified and became a potential danger. On September 1, it made landfall in the Bahamas in Elbow Cay. On September 2, the hurricane remained nearly stationary over the Bahamas as a Category 5 storm. On September 3, Dorian began weakening in intensity as it started moving northwestward, parallel to the east coast of Florida. The hurricane turned to the northeast the next day and made landfall on Cape Hatteras with a Category 1 intensity on September 6. It then transitioned into an extra-tropical cyclone and struck Nova Scotia and then Newfoundland with hurricane-force winds on September 8. Finally it dissipated near Greenland on September 10. Hurricane Dorian caused 63 direct and 7 indirect fatalities. It caused USD 8.28 billion worth of damage. Affected areas included Lesser Antilles, Puerto Rico, The Bahamas, Eastern United States, Eastern Canada, southern Greenland, and Iceland\footnote{\url{https://en.wikipedia.org/wiki/Hurricane_Dorian}}.


\subsection{Community Emergency Response Team Deployment}



In the United States, Community Emergency Response Teams (CERTs) offer a consistent, nationwide approach to volunteer training and organization that professional responders can rely on during disaster situations\footnote{\url{https://www.ready.gov/cert}}. When called upon, CERTs can assist formal humanitarian organizations in a range of disaster response and management tasks. Some CERTs have expanded their team capabilities to provide virtual assistance that includes social media analysis. Montgomery County, Maryland CERT applies a methodological framework as described by \parencite{peterson2019when} when searching for mission-specific content extracted from Twitter. This includes, but is not limited to, performing the following tasks to find reports of damage:

\begin{enumerate}
    \item Use hashtags and keywords to manually search for relevant tweets, including tweets containing images showing some degree of damage.
    \item Analyze tweet text for pertinent cues that would qualify it as valuable. (e.g., context, location, user profile, etc.).
    \item Download damage images into a team collaborative working document and determine the applicability of each image to the mission assignment.
    \item Send summary-of-findings report, including appropriate images, to respective stakeholder (e.g., FEMA).
    \item Repeat above steps throughout operational period.
\end{enumerate}

The above-described methodological framework is effective for social media analysis during disasters, when the mission assignment is focused on text. For example, searching tweets for information indicating road conditions within a region impacted by the disaster. Most social media management tools that Montgomery County, Maryland CERT has used lack the capability to retrieve only tweets containing disaster images. This could hinder future mission assignments related to retrieving visual data because of complex and time-consuming manual steps. For example, first, each tweet would need to be individually checked by a human to determine if an image was included. Secondly, if the tweet did contain an image, and that image was determined to be of value to the mission assignment, it would need to be extracted and placed within a collaborative document. Then potentially another human would determine the applicability of the image to the mission assignment.



Manual analysis of a high-volume data source such as Twitter often leads to information overload~\parencite{hiltz2013dealing}. Therefore, instead of following the above manual steps, we used an automatic Twitter image collection and processing system to find reports of damages caused by Hurricane Dorian as it was progressing. Next, we describe the details of the automatic processing system.

\subsection{Automatic Image Processing System Deployment}
We used AIDR image processing system~\parencite{nguyen2017automatic,imran2014aidr} to start collecting tweets related to Hurricane Dorian on August 30, 2019. The collection ran for about two weeks and stopped on September 14, 2019. In total, approximately 6,890,106 tweets were collected. The below listed keywords were used to collect English language tweets. 

\begin{center}
\begin{tabular}{ m{39em} } 
\textbf{Keywords used to collect tweets} \\
\hline
\textit{HurricaneDorian, Dorian, DorianAlert, Alerts\_Dorian, PuertoRico, DorianMissing, DorianDeaths, HurricaneDorianMissing, HurricaneDorianDeaths, Dorian Missing, Hurricane Dorian Missing, Dorian Found, DorianFound}\\ 
\hline
\end{tabular}
\end{center}

\subsection{Images Processing Modules}
The system has a number of different image analysis modules to process images on Twitter. In this system deployment, we used four of them, which are described below.

{\bf \textit{Image URL deduplication}:} Due to the high number of retweets, some images are re-shared on Twitter thousands of times. Downloading duplicate or near-duplicate images is time-consuming and not helpful for decision-makers. This module keeps track of image URLs and maintains a hash of unique ones. Upon receiving a new image URL, the system determines whether it is unique or not by querying the hash with a time-complexity of O(1)\footnote{\url{https://en.wikipedia.org/wiki/Time_complexity}} i.e., the search takes constant time irrespective of the hash queue length.

{\bf \textit{Image deduplication}:} Images downloaded using unique URLs are not warranted to be actually unique. Different URLs can point to the same image hosted and shared by different web hosts. Moreover, an image could be cropped, resized, or re-shared with additional text inserted on the existing image. Therefore, determining whether an image is a duplicate by comparing it to all the existing images collected by the system to date is crucial. The image deduplication module performs this check by measuring the distance between a newly collected image and existing images using the Euclidean distance on features extracted from images. More specifically, the system uses a deep neural network to extract features from an image and keeps it in a hash. We use a fine-tuned VGG16 model~\parencite{simonyan2014very} and extract features from its penultimate fully-connected (i.e., ``fc2'') layer. A Euclidean distance less than 20 between the features of two images is considered as the two images are duplicate or near-duplicate. Determining an optimal distance threshold is an empirical question, which is not the focus of this work. However, a distance of 20 worked best for our setting.

{\bf \textit{Junk filtering}:} Generally, Twitter is full of noisy content and disasters are not an exception. Research studies have found images of cartoons, advertisements, celebrities, and explicit content shared in tweets related to a disaster event~\parencite{alam2018crisismmd,alam2018SocialMedia}. Trending hashtags are often exploited for this purpose. Such irrelevant content must not be shown to decision-makers during disaster response and recovery efforts given their time is valuable and limited. Unnecessary disruptions must be avoided. The junk filtering module tries to detect irrelevant images by using a deep learning model which is trained to detect irrelevant concepts such as cartoon, celebrities, banners, advertisements. The F1 score (i.e., the harmonic mean of the precision and recall) of this model is 98\%~\parencite{nguyen2017automatic}.

{\bf \textit{Damage severity assessment}:} A unique and potentially relevant image is then finally analyzed by the damage severity assessment module, which determines the level of damage shown in the image. We used a transfer learning technique to fine-tune an existing VGG16 model originally trained on the ImageNet dataset. 
The fine-tuning of the network (all layers) is performed based on the damage-related labeled dataset consisting of three classes. The three classes are \textit{severe}, \textit{mild}, and \textit{none}. The \textit{severe damage} class contains images that show fully destroyed houses, building, bridges, etc. The \textit{mild damage} class contains images that show partially destroyed scenes of houses, building, or transportation infrastructure. The F1 score of this model is 83\%. 

\subsection{Human-in-the-loop for Image Labeling in Real Time}

Automatic systems are not perfect and may make mistakes. It is essential to have some human involvement either to verify the produced results or provide supervision to the system if/when needed~\parencite{imran2014coordinating}. Our system uses human-in-the-loop for both verification and gaining supervision purposes. Data items processed by the system are used to take samples for humans to verify and guide the system if a mistake is identified. Such mistakes could be false positives or false negatives. Human-labeled items would then be ideally fed back to the system for retraining a new model for enhanced performance. 

\begin{figure}
\centering
\includegraphics[width=0.8\linewidth]{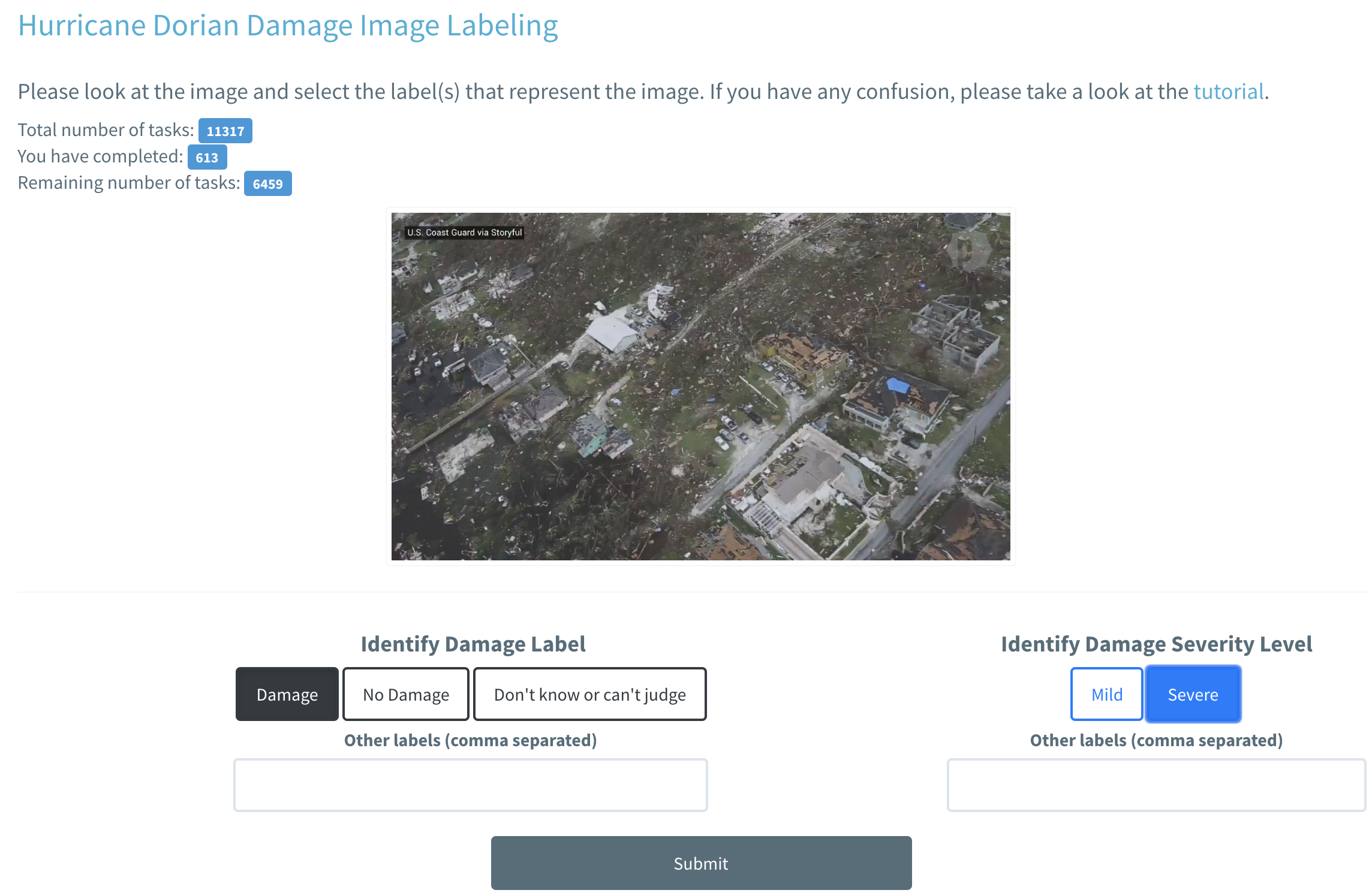}
\caption{Web-based interface for human assessors to verify system predictions and relabel images if required. The highlighted labels are system predicted.}
\label{fig:mm_image_labeling}
\end{figure}

To involve humans in the verification and supervision process, we used our MicroMappers crowdsourcing system\footnote{\url{https://micromappers.qcri.org/}}. Images downloaded and classified by our data processing system were first used to take samples. We performed this sampling every couple of hours during the operational period for Montgomery County, Maryland CERT (details in the next section). In most of the samples, we selected all \textit{severe damage} and \textit{mild damage} images and some from the \textit{none} class from the system-processed images in a given time-window of past $T$-hours. We did not fix the $T$ value, i.e., number of hours, as human processing speed depends on many unknown factors. The sampled images were then shown to human experts. For this deployment, we decided to only crowdsource the output of our damage severity assessment module, which classifies an image into one of three damage levels (i.e., classes), as described above. On a web interface, we showed an image along with the system predicted class to the expert. The human expert either agreed or disagreed with the machine classification. In the case where they disagreed with the machine classification, they would provide a new label to the image. Figure~\ref{fig:mm_image_labeling} depicts the crowdsourcing interface. The interface first showed the options (\textit{Damage}, \textit{No Damage}, and \textit{Don't know or can't judge}), which can be seen on the left. If the human selected the \textit{Damage} label, the interface would further show two severity levels (\textit{Mild}, \textit{Severe}), which would appear on the right side of the screen. The human would select one of these two severity labels and submit their assessment. If the human selected \textit{``Don't know or can't judge"}, then the system would not show the two additional severity labels. The human experts were allowed to provide additional comments using the text boxes on the interface.

\begin{figure}
\centering
\includegraphics[width=0.8\linewidth]{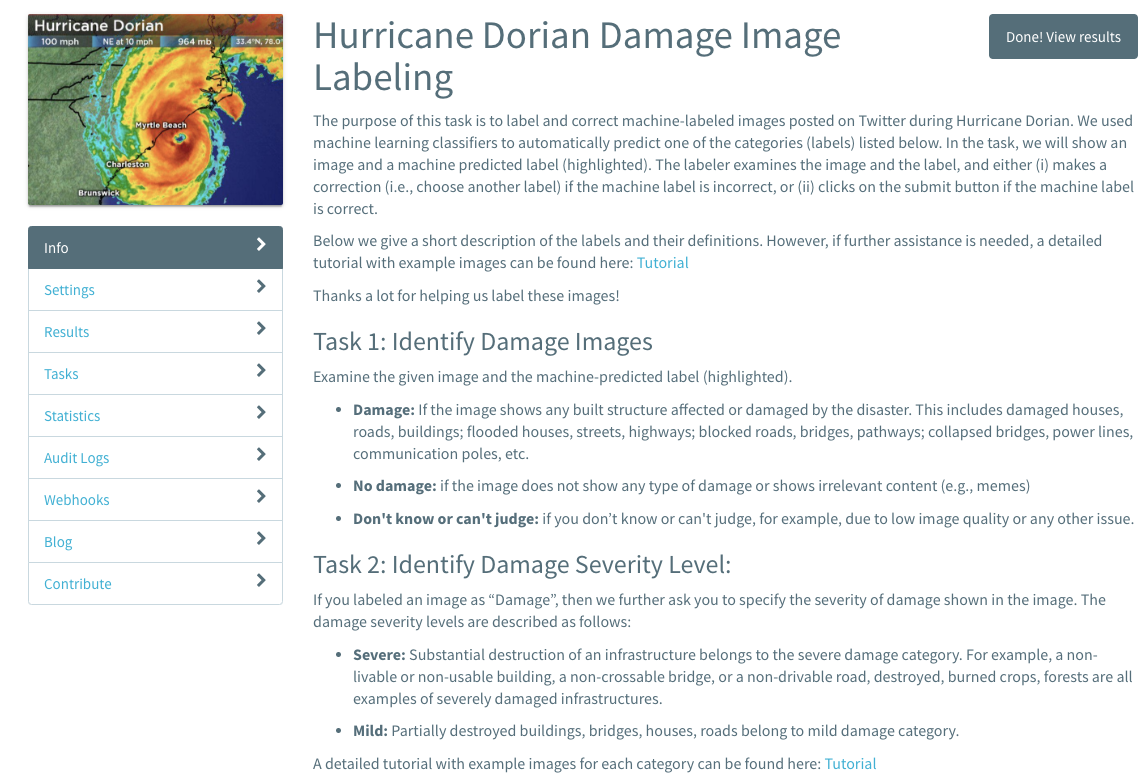}
\caption{Task description page showing details of the tasks including classes definition}
\label{fig:mm_task_details}
\end{figure}

In addition to the labeling interface, we established two other pages, one for showing the task details (Figure~\ref{fig:mm_task_details}) and the second for a detailed tutorial\footnote{\url{https://ibb.co/DztXbTy}} with concrete examples for each class. Each human expert was instructed to go through the tutorial before beginning their labeling effort. 

%% file: results.tex
\section{Data and Results}
\label{sec:data_results}








\subsection{Data Statistics}
As shown in Table~\ref{tbl:data_stats}, out of all 6,890,106 tweets collected, 280,063 unique image URLs were found. The total number of downloaded images was 279,819. Around 244 images failed to download due to one of several reasons, e.g., the tweet author deleted the actual tweet, the image host server was down, or connection timed out, etc. 

\begin{table}[h!]
\caption{Hurricane Dorian tweets and image data statistics}
\centering
\begin{tabular}{c c c c c} 
 \hline
 Total tweets & Unique image URLs & Downloaded images & Failed to download \\ 
 \hline
 6,890,106 & 280,063 & 279,819 & 244 \\ 
 \hline
\end{tabular}

\label{tbl:data_stats}
\end{table}

\subsection{Automatic Classification Results}
The 279,819 images, which were successfully downloaded, were then analyzed by the image processing modules described in the previous section. An image-based deduplication was performed as the first step followed by the image relevancy check executed by the \textit{Junk filtering} module. All relevant images were then ingested by the \textit{Damage Severity Assessment} module to determine if they contained any damage. If the image contained damage, it was classified according to the severity shown in the scene.  


\begin{table}[h!]
\caption{Image-based automatic processing: Results of unique, relevant, and damage images.}
\centering
\begin{tabular}{c c c c c} 
 \hline
 Unique images & Relevant images  & Images with damage & Severe damage & Mild damage\\ 
 \hline
 119,767 & 77,580 & 26,386 & 11,044 & 15,342 \\ 
 \hline
\end{tabular}

\label{tbl:image_stats_good}
\end{table}

Table~\ref{tbl:image_stats_good} shows the number of images which were found as unique, relevant, and with some level of damage -- specifically severe and mild damage. Out of 279,819 images, the image-based deduplication module found 119,767 unique images, which was around 42\% of the whole set. As described earlier, this image-based deduplication module relies on deep features extracted from images using a deep neural network. Due to the high retweet/re-sharing ratio on Twitter, even during a large-scale natural disaster, 58\% of the images were identified as exact or near-duplicate by the system. At this stage, the process of automatically finding near-duplicate images had already reduced the chance of information overload affecting the human experts.

\begin{figure}[h!]%
    
    \centering
    \subfloat{{\includegraphics[width=3.2cm]{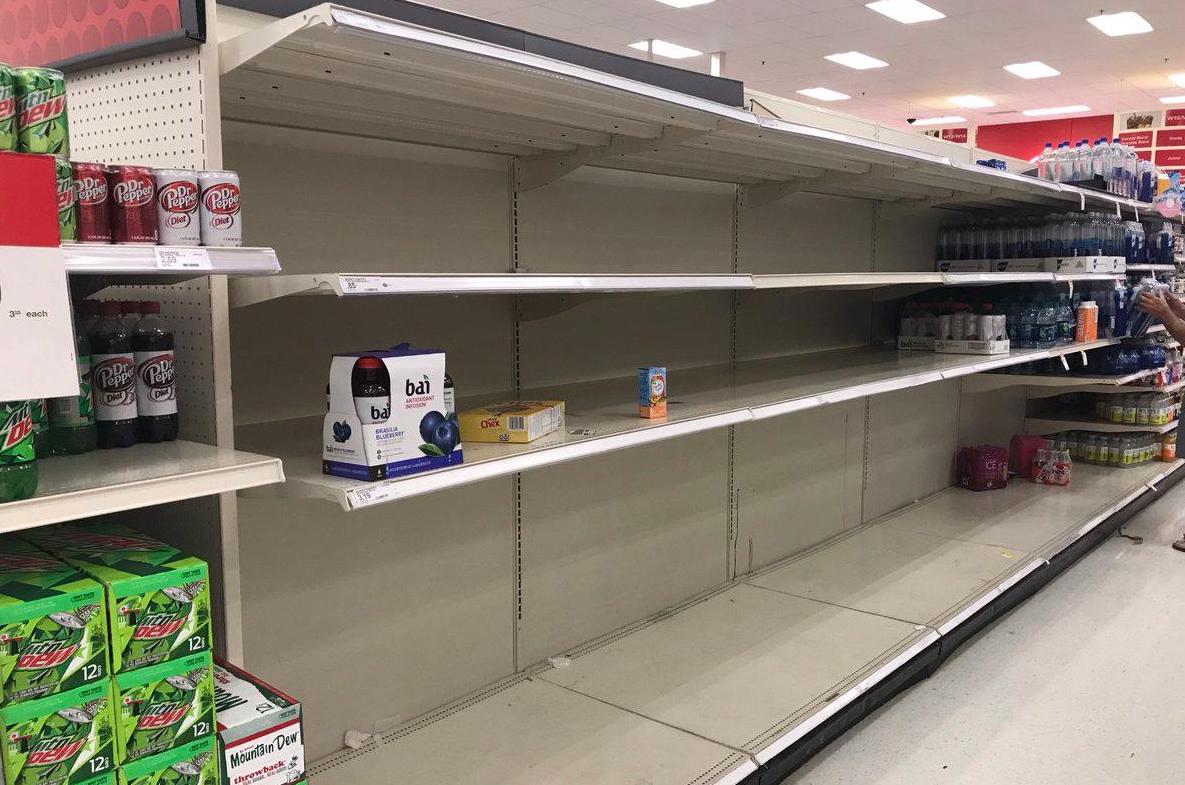}}}%
    \qquad
    \subfloat{{\includegraphics[width=3.2cm]{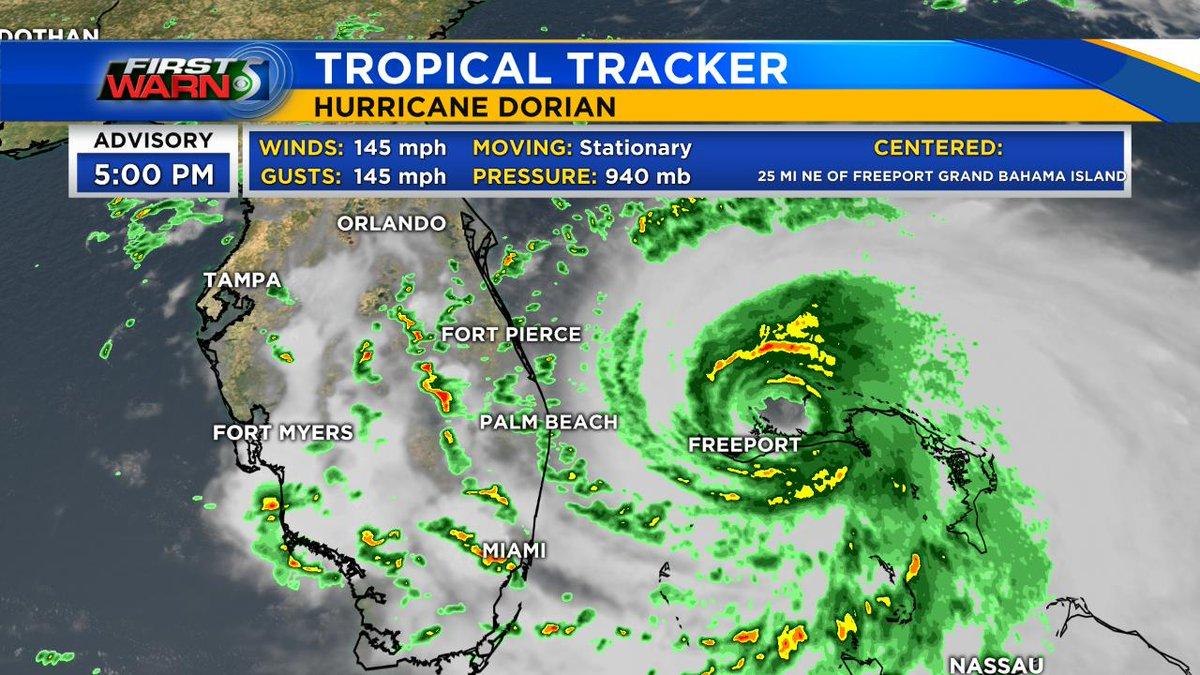} }}%
    \qquad
    \subfloat{{\includegraphics[width=3.2cm]{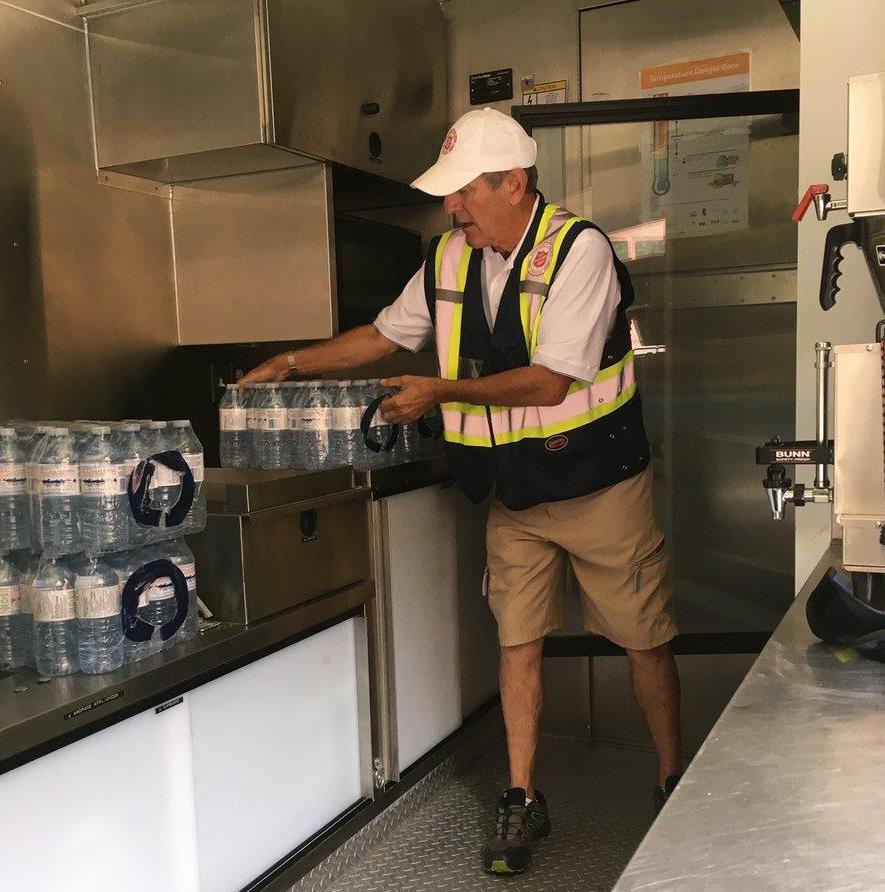} }}%
    \qquad
    \subfloat{{\includegraphics[width=3.2cm]{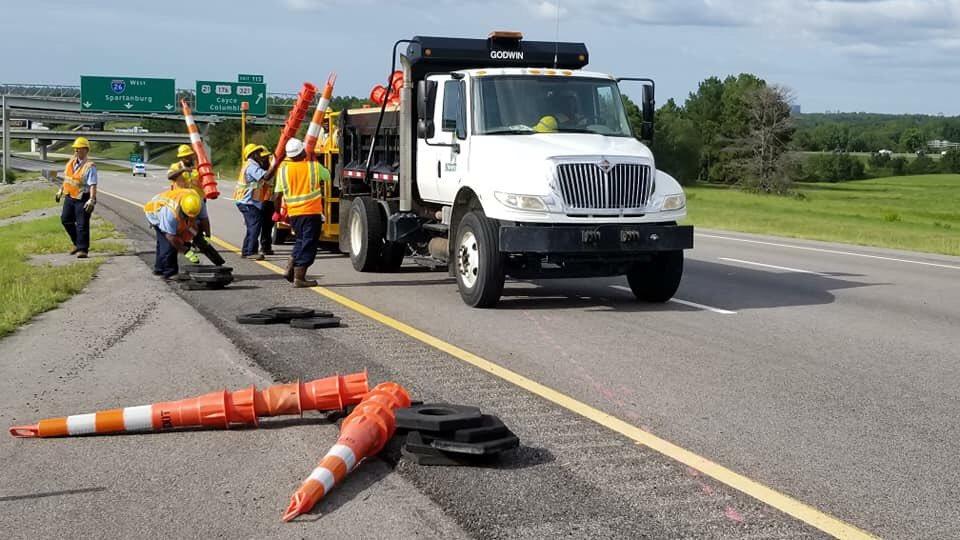} }}%
    \caption{Images that are relevant but do not show any damage}%
    \label{fig:relevant_no_damage_imgs}%
    
\end{figure}

\begin{figure}[h!]
\centering
\includegraphics[width=1\linewidth]{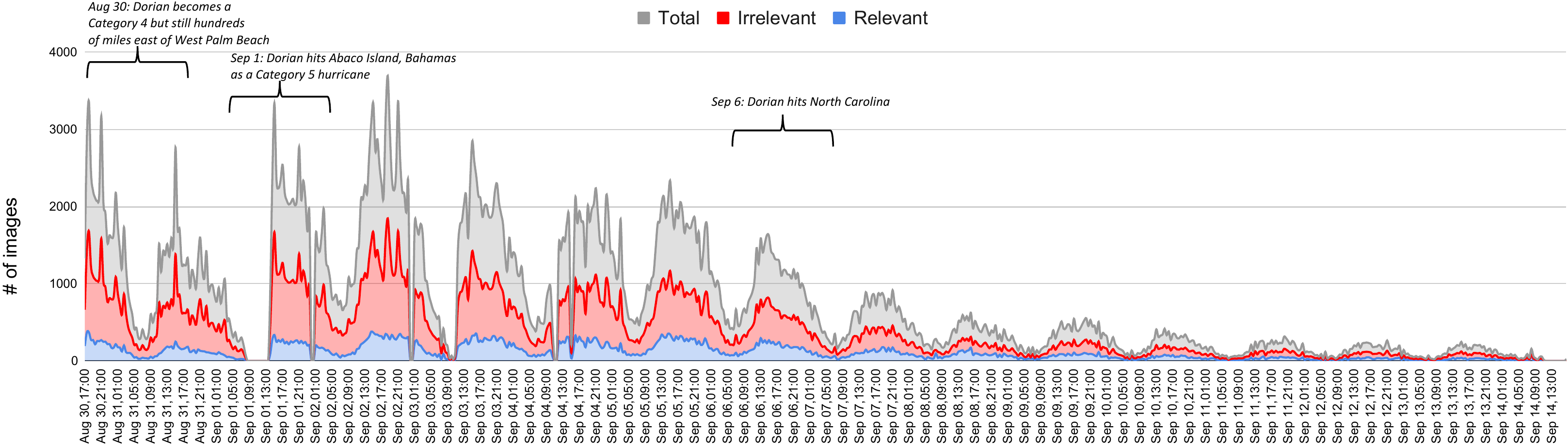}
\caption{Distribution of total, relevant, and irrelevant images over the duration of the event}
\label{fig:dist_relevant_irrelevant}
\end{figure}

Furthermore, out of the 279,819 images, 77,580 
were identified as relevant by the system. These images did not contain cartoons, celebrities, banners, advertisements, etc. Among the relevant images, some contained damage scenes while others did not. Figure~\ref{fig:relevant_no_damage_imgs} shows a few images that did not show any damage but were identified as relevant. Many of the relevant images showed hurricane maps or some other scene associated to rescue efforts. We show the distribution of total, relevant, and irrelevant images for the whole deployment period in Figure~\ref{fig:dist_relevant_irrelevant}.

\begin{figure}[h!]%
    \centering
    \subfloat{{\includegraphics[width=5cm]{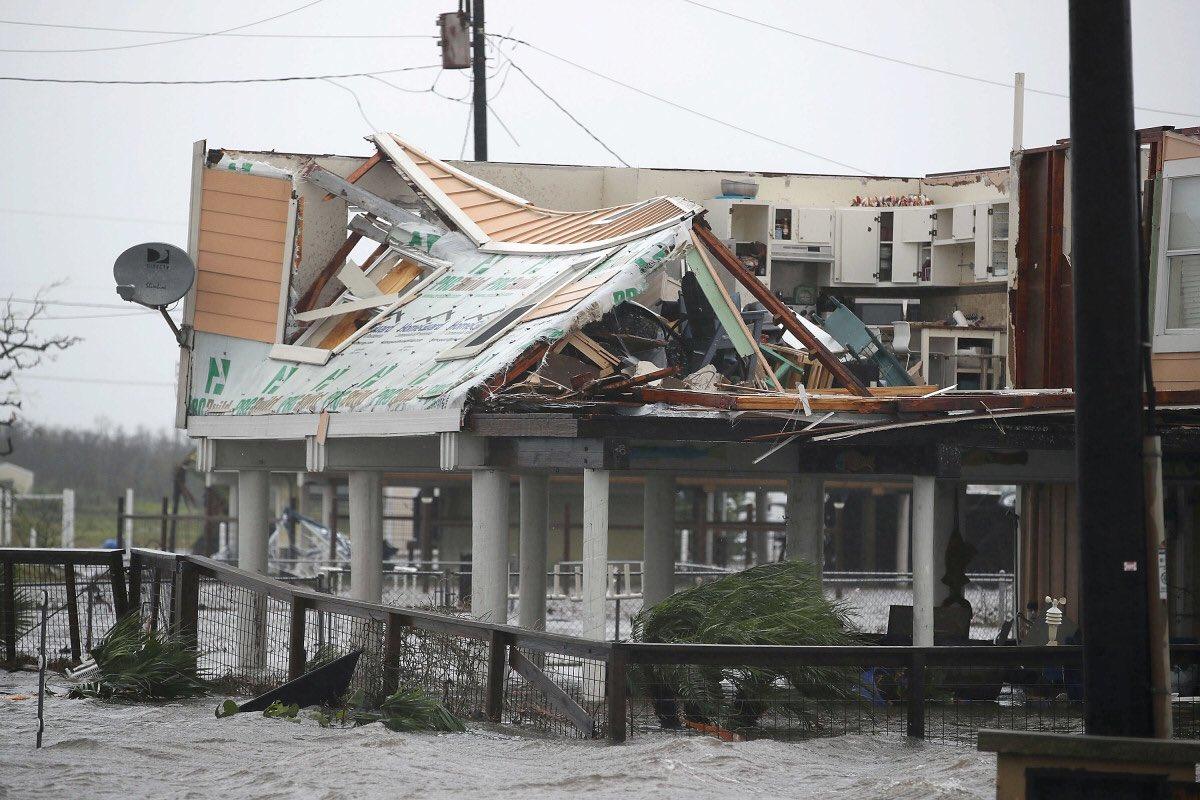}}}%
    \qquad
    \subfloat{{\includegraphics[width=5cm]{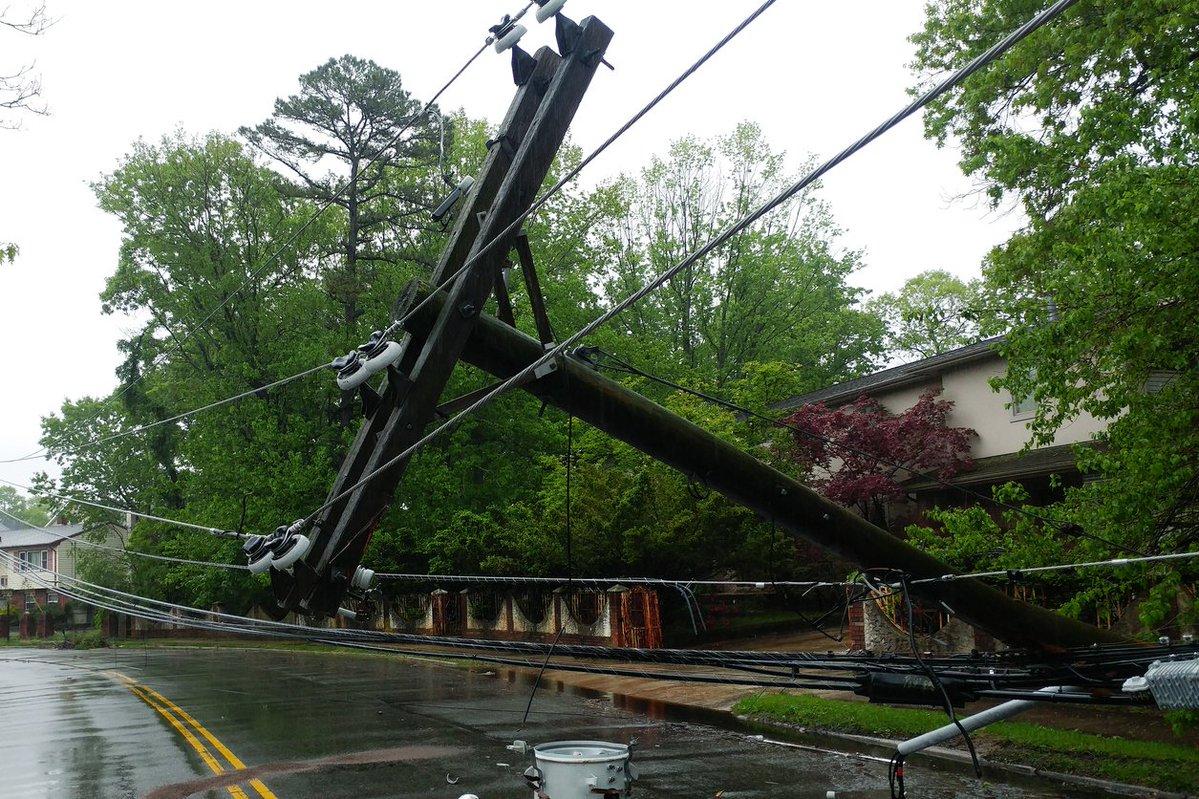} }}%
    \qquad
    \subfloat{{\includegraphics[width=5cm]{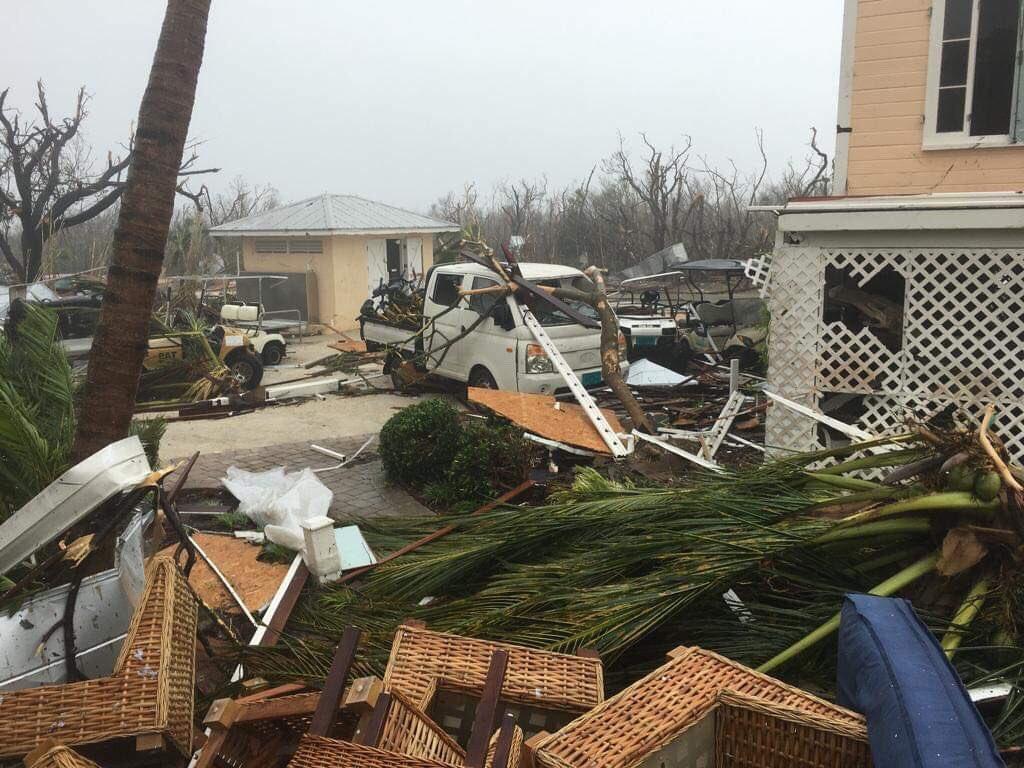} }}%
    \qquad
    \subfloat{{\includegraphics[width=5cm]{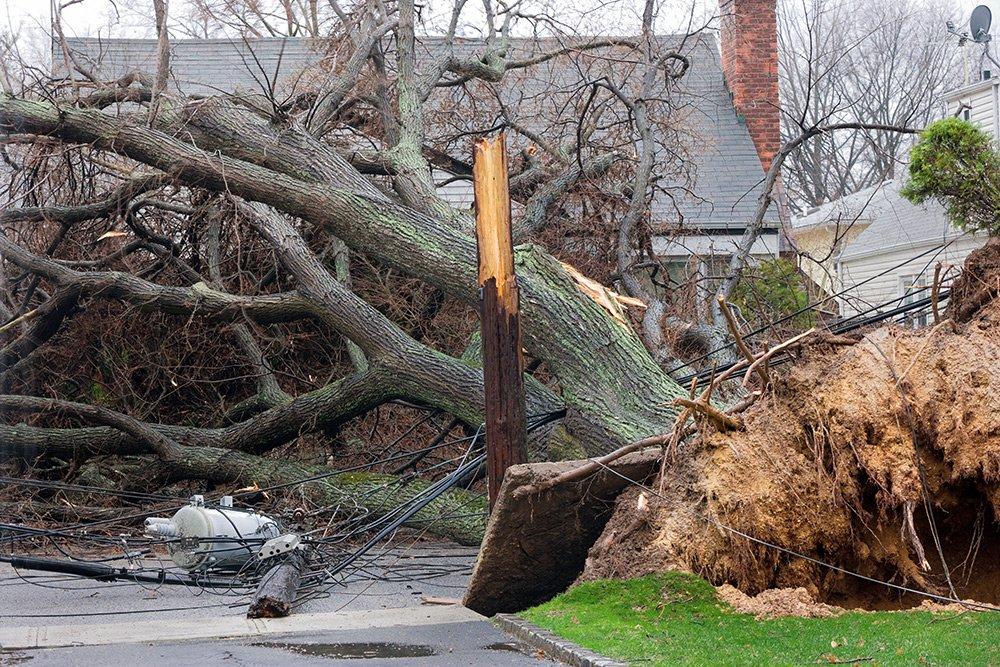} }}%
    \caption{Severe damage images identified by the automatic system}%
    \label{fig:severe_imgs}%
\end{figure}

\begin{figure}[h!]%
    \centering
    \subfloat{{\includegraphics[width=5cm]{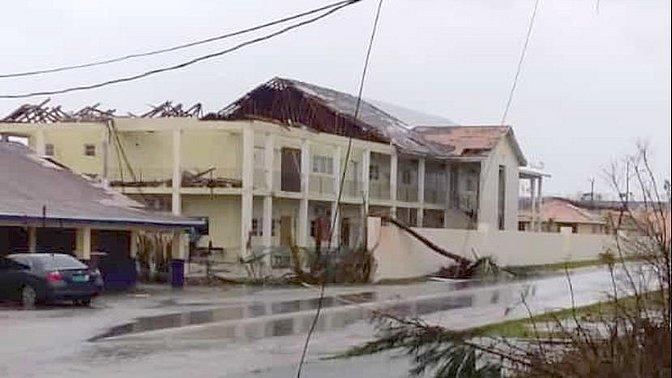}}}%
    \qquad
    \subfloat{{\includegraphics[width=5cm]{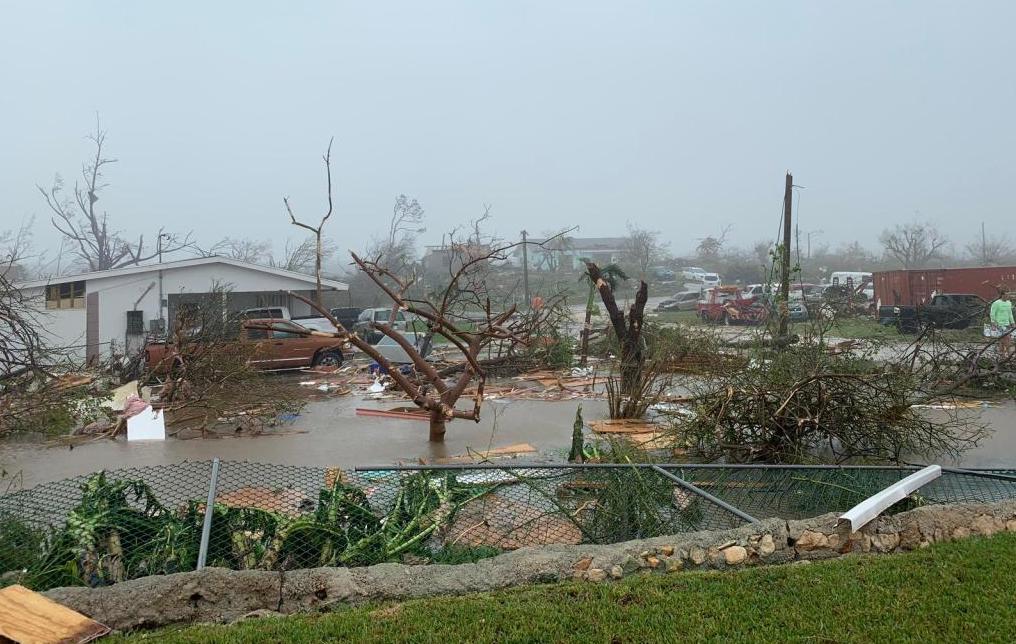} }}%
    \qquad
    \subfloat{{\includegraphics[width=5cm]{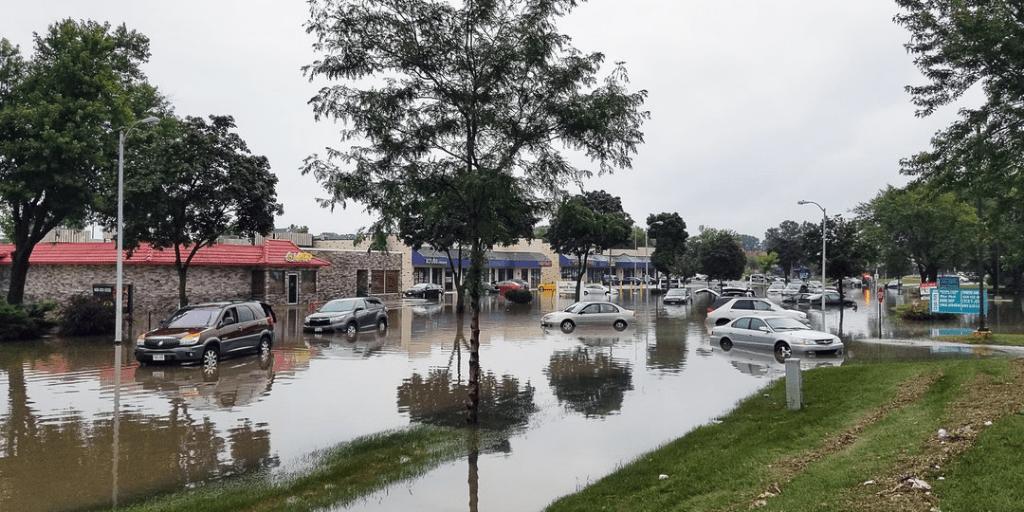} }}%
    \qquad
    \subfloat{{\includegraphics[width=5cm]{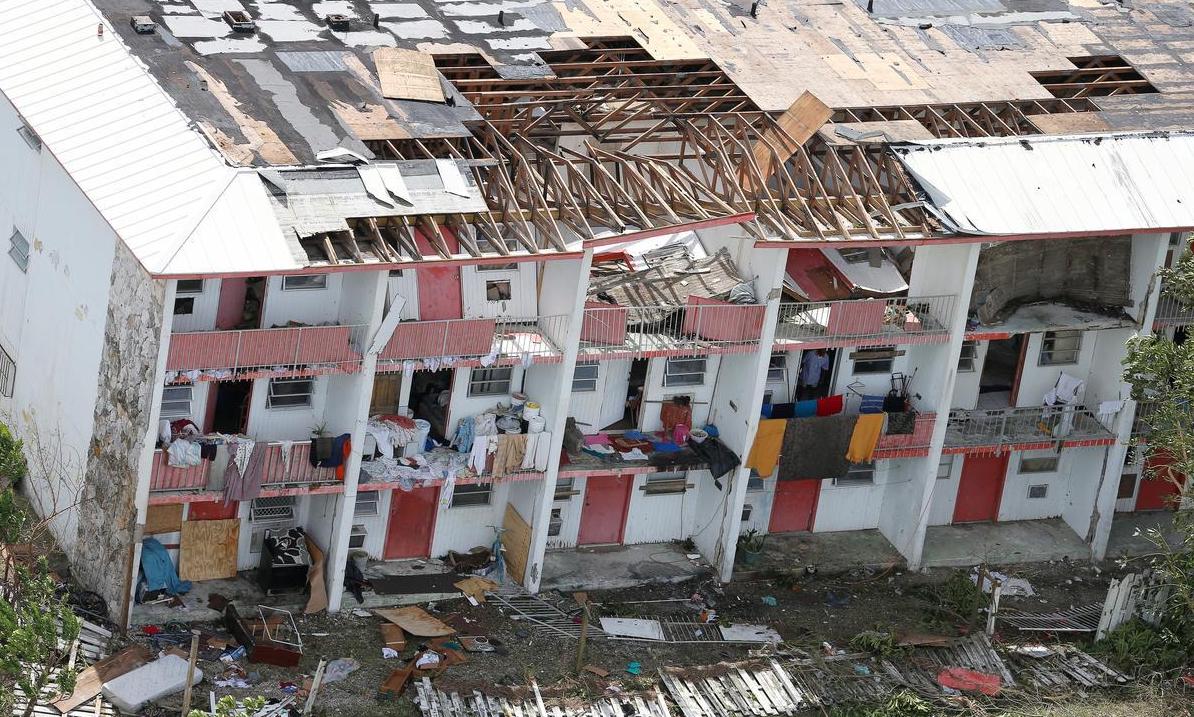} }}%
    \caption{Mild damage images identified by the automatic system}%
    \label{fig:mild_imgs}%
\end{figure}

\begin{figure}[h!]
\centering
\includegraphics[width=1\linewidth]{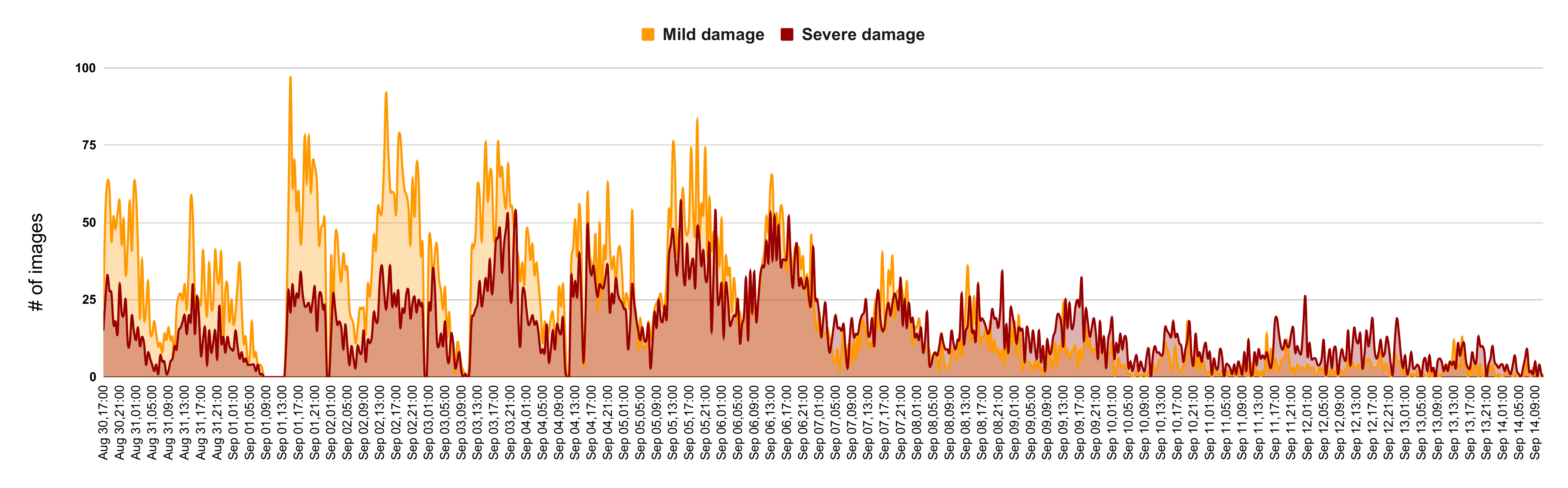}
\caption{Distribution of severe and mild damage images over the duration of the event}
\label{fig:dist_severe_mild}
\end{figure}

Out of all relevant images, 26,386 were identified as containing some damage where 11,044 showed severe and 15,342 showed mild damage. The images with damage scenes were around ${\sim}$10\% of all the downloaded images. The system's ability to filter out ${\sim}$90\% of images as potentially not containing any damage content is a significant reduction in risk of information overload to humans. Figure~\ref{fig:severe_imgs} contains a few images, which according to the system showed severe damage. Figure~\ref{fig:mild_imgs} shows a few images, which according to the system included mild damage. We show the distribution of mild and severe damage images as classified by the system for the whole duration period in Figure~\ref{fig:dist_severe_mild}.


Finally, Table~\ref{tbl:image_stats_bad} shows distribution of images identified as duplicate, not relevant, and containing no damage. 

\begin{table}[h!]
\caption{Image-based automatic processing: Results of duplicate, not relevant, and no damage images}
\centering
\begin{tabular}{c c c c c c} 
 \hline
 Duplicate images & Not relevant images & Images with no damage \\ 
 \hline
 160,052 & 202,239 & 253,433 \\ 
 \hline
\end{tabular}

\label{tbl:image_stats_bad}
\end{table}

\subsection{Human Verification and Image Labeling Results}
As Hurricane Dorian progressed, human experts from Montgomery County, Maryland CERT (N=28) were asked to look at an evolving sample of system processed images to verify if the system was producing desired results. The experts were also instructed to correct any identified mistakes done by the system. Given they were trained domain experts, not employed from an online paid crowdsourcing platform, we trusted their judgements without the need to ask multiple assessors. This meant each image was assessed by only one human expert. At the conclusion of the CERT operational period, their team lead reviewed around ${\sim}$2K of the completed tasks for quality assurance. This feedback can be found in the \textit{Discussion} section.  

\begin{table}[htbp]
\begin{center}
\caption{Damage detection task confusion matrix---system vs. human judgments}
\label{tbl:confusion-matrix_task_damage}
\begin{tabular}{|c|c|c|c|}\hline
\multicolumn{2}{|c|}{ \multirowcell{2.2}{N=28,050}}
& \multicolumn{2}{c|}{\bfseries Machine} \\%
  \cline{3-4}
   \multicolumn{2}{|c|}{} & \thead{~~~Damage~~~} & \thead{No Damage}\\%
  \hline
  \multirowcell{2.7}{\bfseries Human}
    & \thead{Damage} & \cellcolor{blue!20}2,088 & 712\\%
    \cline{2-4}
    & \thead{No Damage} & 5,954 & \cellcolor{blue!20}19,296\\%
    \hline
\end{tabular}
\end{center}
\end{table}

Table~\ref{tbl:confusion-matrix_task_damage} shows the results of the damage detection task. In total, the human experts analyzed 29,136 
images over a 42-hour operational period from 8:00pm on September 6 to 2:00pm on September 8. These images were initially processed by the system and contained scenes of both damage and no damage. Moreover, when an image contained damage, it had one of three damage severity labels (severe, mild, none) assigned by the system. Of all 29,136 analyzed images, 1,086 were labeled as ``Don't know or can't judge'' by the experts. This could have been due to several reasons including blurred/low quality images, closeup shots, too dark/small, or an image containing text. From the remaining set (i.e., 28,050), the experts agreed with the system predictions for 21,384 images. This agreement can be seen in the two diagonal colored cells of the Table~\ref{tbl:confusion-matrix_task_damage}, where in 2,088 cases both system and human agreed that the image showed some damage and 19,296 cases the image showed no damage. However, there were 6,666 (5,954 + 712) images which the experts did not agree with the system. Based on the results of this human analysis, we compute the system accuracy = 76\% .

\begin{table}[htbp]
\begin{center}
\caption{Damage severity assessment task confusion matrix---system vs. human judgments}
\label{tbl:confusion-matrix_task_severity}
\begin{tabular}{|c|c|c|c|c|}\hline
\multicolumn{2}{|c|}{ \multirowcell{2.2}{N=28,050}}
& \multicolumn{3}{c|}{\bfseries Machine} \\%
  \cline{3-5}
   \multicolumn{2}{|c|}{} & \thead{Severe Damage} & \thead{Mild Damage} & \thead{~~~~ None ~~~~}\\
  \hline
  \multirowcell{3.7}{\bfseries Human}
    & \thead{Severe Damage} & \cellcolor{blue!20}710 & 384 & 357\\%
    \cline{2-5}
    & \thead{Mild Damage} & 113 & \cellcolor{blue!20}881 & 355\\%
    \cline{2-5}
    & \thead{None} & 721 & 5,233 & \cellcolor{blue!20}19,296\\%
    \hline
\end{tabular}
\end{center}
\end{table}

For the second task, which aimed to assess the severity of damage in an image, the results are shown in Table~\ref{tbl:confusion-matrix_task_severity}. The human experts agreed with the system 20,887 times, as shown in the three diagonal colored cells. However, we received a disagreement for 7,163 images. Based on the results of this human analysis, we measured the system accuracy as 74\% for this task. 

We report detailed system performance results in terms of precision, recall, F1, and accuracy for both tasks in Table~\ref{tbl:system_performance_both_tasks}. The system achieved a precision of 0.89 for both tasks, which is a reasonable score. However, the recall scores are a little lower, i.e., 0.76 for task 1, and 0.74 for task 2. 

\begin{table}[h!]
\centering
\caption{System performance for both tasks}
\label{tbl:system_performance_both_tasks}
\begin{tabular}{l c c c c} 
 \hline
Classification tasks  & Accuracy & Precision & Recall & F1 \\ 
 \hline
 Task 1: Damage; No damage & 0.76 & 0.89 & 0.76 & 0.80 \\ 
 Task 2: Severe; Mild; None & 0.74 & 0.89 & 0.74 & 0.80 \\ 
 \hline
\end{tabular}
\end{table}

%% file: discussion_conclusion.tex
\section{Discussion and Error Analysis}
\label{sec:discussion}

 

 

 From an emergency manager's point of view, it is important the system does not miss any damage reports, regardless of the severity of damage. Missed damage reports could provide relevant information on an impacted area that had minimal or no actionable intelligence immediately available for decision-making. Therefore, among other cases, `false negative' are most important for us to analyze. For example, when the machine predicts an image as not containing any damage i.e., \textit{``None''} but the human expert labels it as \textit{``Severe''} or \textit{``Mild''}. There were 357 cases where Machine=None \& Human=Severe and 355 cases where Machine=None \& Human=Mild. These cases can be seen in Table~\ref{tbl:confusion-matrix_task_severity} and are analyzed next. 
 
\begin{figure}[h!]
\centering
\includegraphics[width=1\linewidth]{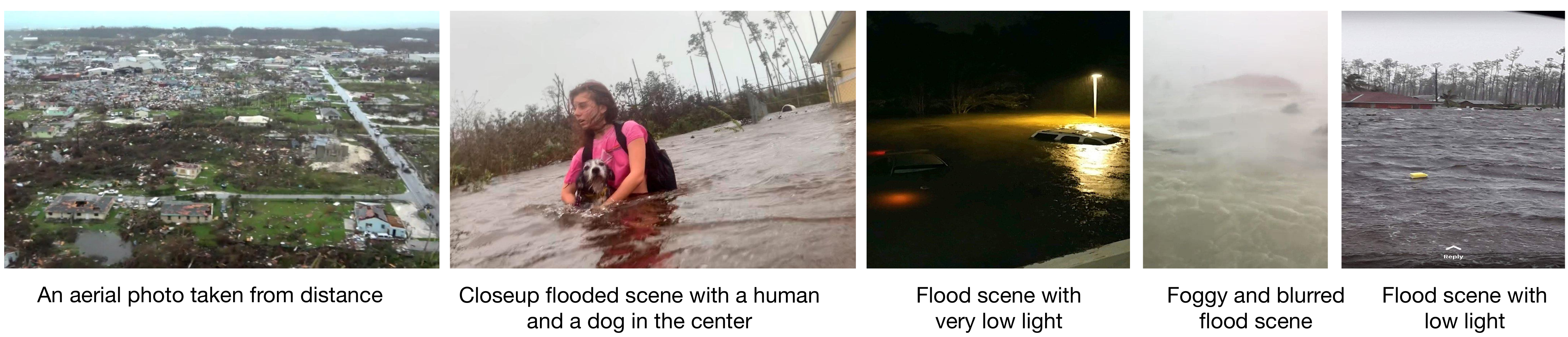}
\caption{False negative examples where Machine:None \& Human:Severe}
\label{fig:mn_hs}
\end{figure}

 {\bf Machine:None vs. Human:Severe:} Figure~\ref{fig:mn_hs} shows a few images where the machine prediction was \textit{None} (i.e., no damage) and human assessment was \textit{Severe damage}. Our in-depth analysis of these 357 images reveals that in most of these false-negative cases, the machine mainly missed flooded scenes. Another main pattern that emerged is where images with low light confused the machine such as the third image from the left in Figure~\ref{fig:mn_hs}. Also, aerial images covering a wide area caused issues for the machine to understand them (i.e., first image on the left). Image collages are also a source of problem for the machine. We define an image collage as multiple images joined together to appear as one. Such cases create even more challenges to accurately classify damage severity when the level of damage in at least one of the images contradicts the level of damage in another image within the same collage.

\begin{figure}[h!]
\centering
\includegraphics[width=1\linewidth]{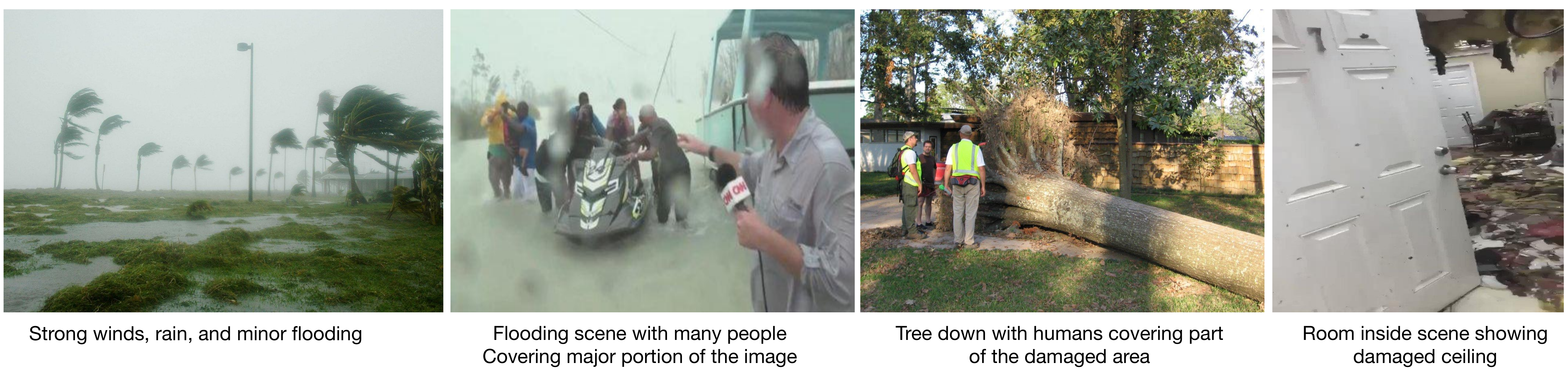}
\caption{False negative examples where Machine:None \& Human:Mild}
\label{fig:mn_hm}
\end{figure}

 {\bf Machine:None vs. Human:Mild:} Figure~\ref{fig:mn_hm} shows a few cases out of 355 where the machine prediction was \textit{None}, but according to the human experts these images showed \textit{Mild damage}. Our analysis of these cases revealed that most of the damage appearing in an image was covered by another object. This caused issues for the machine to classify them as a damage image. In the second and third image from the left in Figure~\ref{fig:mn_hm}, it can be seen there are people standing and covering some parts of damage scenes. Whereas in the fourth image, a white door is covering 80\% of the damage scene, leaving only a small area for the machine to predict it as a mild damage case. 
 Moreover, we noticed that scenes with trees showing strong winds were also missed by the machine. 
 
 For the above two cases, further investigation revealed that our damage severity assessment model's training data lacks flooded and strong winds scenes, which is one of the reasons the model missed many such cases. 
 
 While responding to a disaster, time is limited and precious to decision-makers. Having them preoccupied with looking at irrelevant data is not appropriate and risks creating information overload consequences. Therefore, another important area for us to understand are false positives our system generates. As shown in Table~\ref{tbl:confusion-matrix_task_severity}, there were 5,954 (721 + 5,233) images which, according to the machine, either contained \textit{Severe damage} (i.e., 721 cases) or \textit{Mild damage} (i.e., 5,233 cases), but according to the human experts these cases were \textit{None}---meaning they did not show any damage. Next, we study these two cases.
 
\begin{figure}[h!]
\centering
\includegraphics[width=1\linewidth]{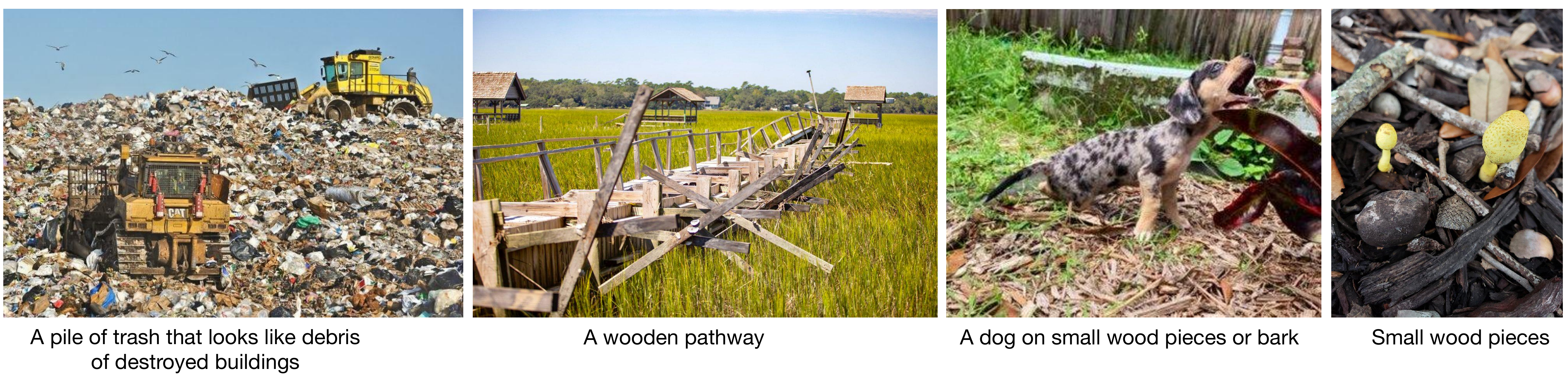}
\caption{False positive examples where Machine:Severe \& Human:None}
\label{fig:ms_hn}
\end{figure}

 {\bf Machine:Severe vs. Human:None:} We extensively analyzed these 721 images. A few of them are shown in Figure~\ref{fig:ms_hn}. Our analysis revealed that most of the images appeared to contain some damage, but actually they did not. Many images contained scenes with irregular arrangements of wooden pieces, which deceived the model to predict them as damage. The first image from the left in Figure~\ref{fig:ms_hn} shows a pile of trash that could be interpreted as debris of a destroyed built infrastructure. The second image has a wooden pathway with irregular arrangements of lumber. Perhaps, part of the pathway is slightly damaged, but it is not a severe damage scene. Similarly, the other two images caused confusion for the model. Having an ability to identify non-damage scenes which resemble damage scenes would be one of the most challenging tasks to address from the machine's modeling point of view. More hard negative examples would help models better understand and discriminate between positive and negative cases.
 
\begin{figure}[h!]
\centering
\includegraphics[width=1\linewidth]{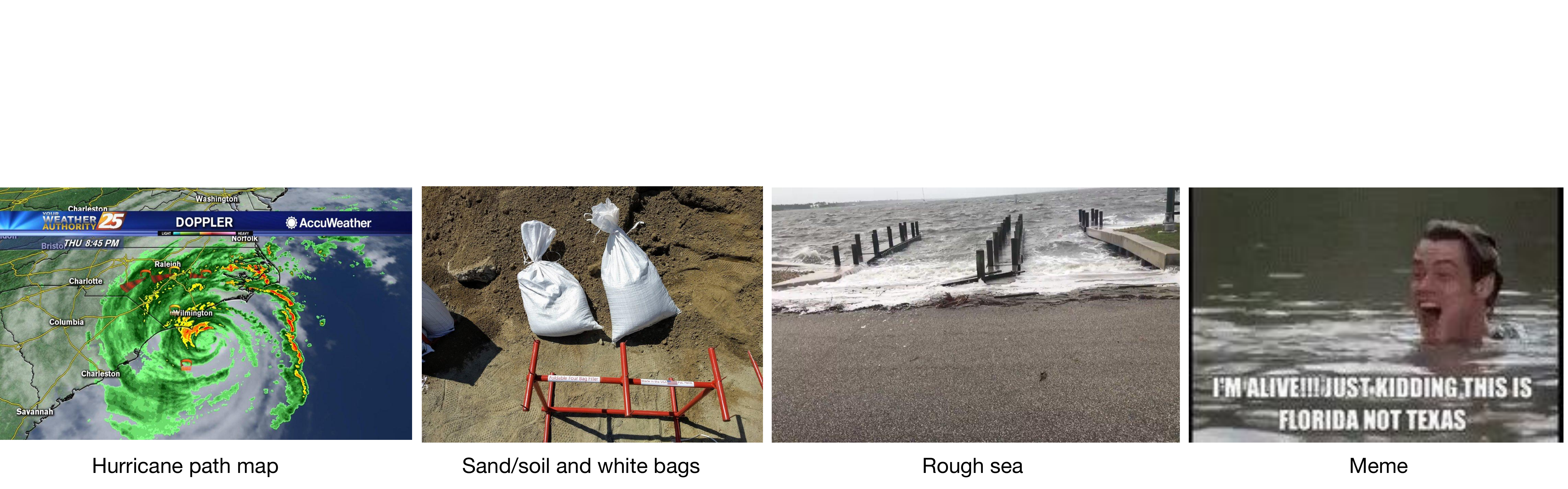}
\caption{False positive examples where Machine:Mild \& Human:None}
\label{fig:mm_hn}
\end{figure}

{\bf Machine:Mild vs. Human:None:} Figure~\ref{fig:mm_hn} shows a few images from this category. Our analysis revealed that the majority of these images showed maps depicting the hurricane's  path as seen in the first image from the left of Figure~\ref{fig:mm_hn}. Among other scenes, there were rough sea images or memes with flooding scenes (i.e., last image from left). Furthermore, we also noticed this category contained many images with people standing in groups or performing some activity. Overall, this category shows more variation in the scenes compared to the other categories. The misclassifications relating to the maps and images where there are people can be easily fixed by feeding more hard negative examples to the machine. However, scenes of rough seas and flooding closely border between the mild or severe categories and thus would be hard to accurately tackle.
 
\subsection{Challenges and Future Work} 
Based on feedback from the human experts, we identified a number of weaknesses and challenges that our image processing models faced. We list these challenges as future work below.

\textbf{$\bullet$ Flood scene variations:} Capturing different variations in flood scenes such as flooding on roads, in houses, forests, or fields is important yet challenging for machine learning models. In our case, this problem occurred mainly due to the lack of appropriate training data that represented such variations. However, in some cases, even a sufficient amount of labeled data might not be enough to resolve ambiguities between a natural scene and a disaster scene. For example, rough sea scenes should not be confused as flood scenes. These difficult cases require additional considerations while training machine learning models and also raise awareness to the need for further research on effective integration of human intelligence into machine learning models.

\textbf{$\bullet$ Low-light damage scenes:} We noticed many foggy and low-light scenes were missed by our models. Similar to the previous challenge, lack of training data collected in low-light conditions caused our models to miss such cases. Addressing this issue is important from a time perspective for decision-makers when a disaster occurs at night time. Accurately classified images can provide awareness of the severity of damage before daylight arrives, thus saving time and allowing for some decisions to begin to be made (e.g., resource allocation planning). In addition to collecting more appropriate training data, other image processing techniques can be used to adjust image contrast, brightness, or saturation as pre-processing steps before feeding them into the model.

\textbf{$\bullet$ Wide-area and aerial images:} Images taken from afar often cover a wide area that shows many objects such as houses, trees, sky, etc. 
These images do not only show objects at a much smaller scale than ground-level images but they also often contain scenes with a mix of both damaged and undamaged objects and areas. Due to such large differences in the scale of objects and areas, it may not be ideal to design a single model that operates on both aerial and ground-level images for any given task.
In particular for the damage assessment task, the ideal solution may require designing separate aerial and ground-level models with more localized (i.e., object-level) damage detection and assessment capabilities.

\textbf{$\bullet$ Maps and memes:} Our models suffered while identifying maps and memes. However, we noticed this deficiency is mainly due to the lack of appropriate labeled data on which our models were initially trained. Adding more suitable training images would help eradicate this problem.

\textbf{$\bullet$ Damage-resembling scenes:} Images that show scenes resembling damaged objects or areas constitutes a big challenge for automatic image processing models. We identified around 700 such cases during our deployment. Machine learning for such scenes may need additional semantic information about objects surrounding a damage scene to help models understand. For example, if a nearby crop field shows intact healthy crops, then it is less likely that overall image shows severe damage.

\section{Conclusions}
\label{sec:conclusions}
Rapid damage assessment provides crucial information about damage severity caused by a disaster in the early stages of response. Humanitarian and formal response organizations rely on field assessment reports, remote sensing methods, or satellite imagery to perform damage assessment. This work leveraged imagery data shared on Twitter to identify reports of damages using image processing techniques based on deep neural networks. Moreover, the image processing system filters out duplicate and irrelevant images which are not useful for decision-makers responding to the disaster. The system was activated before Hurricane Dorian made landfall in the Bahamas and ran for 13 days. Over a 42-hour operational period of collaborating with our partner volunteer response organization, the damage reports identified by the system were examined by the domain experts of this organization, whose feedback revealed that the system achieved an accuracy of 74\% and 76\% for the two damage assessment tasks. Although these scores show the system's effectiveness to process real-world disaster data, we identified a number of shortcomings of our machine learning models, which are listed in the previous section and considered as potential future work.

%% file: main.bib
@inproceedings{erdelj2016uav,
  title={UAV-assisted disaster management: Applications and open issues},
  author={Erdelj, Milan and Natalizio, Enrico},
  booktitle={2016 international conference on computing, networking and communications (ICNC)},
  pages={1--5},
  year={2016},
  organization={IEEE}
}

@article{Mouzannar2018,
author = {Mouzannar, Hussein and Rizk, Yara and Awad, Mariette},
journal = {15th International Conference on Information Systems for Crisis Response and Management (ISCRAM 2018)},
number = {May},
pages = {529--543},
title = {{Damage Identification in Social Media Posts using Multimodal Deep Learning}},
year = {2018}
}

@article{alam2018SocialMedia,
author = {Firoj Alam and Ferda Ofli and Muhammad Imran},
title = {Processing Social Media Images by Combining Human and Machine Computing during Crises},
journal = {International Journal of Human–Computer Interaction},
volume = {34},
number = {4},
pages = {311-327},
year  = {2018},
publisher = {Taylor & Francis},
doi = {10.1080/10447318.2018.1427831},
URL = { 
        https://doi.org/10.1080/10447318.2018.1427831
},
eprint = { 
        https://doi.org/10.1080/10447318.2018.1427831
}
}

@article{kryvasheyeu2016rapid,
  title={Rapid assessment of disaster damage using social media activity},
  author={Kryvasheyeu, Yury and Chen, Haohui and Obradovich, Nick and Moro, Esteban and Van Hentenryck, Pascal and Fowler, James and Cebrian, Manuel},
  journal={Science advances},
  volume={2},
  number={3},
  pages={e1500779},
  year={2016},
  publisher={American Association for the Advancement of Science}
}

@article{barrington2012crowdsourcing,
  title={Crowdsourcing earthquake damage assessment using remote sensing imagery},
  author={Barrington, Luke and Ghosh, Shubharoop and Greene, Marjorie and Har-Noy, Shay and Berger, Jay and Gill, Stuart and Lin, Albert Yu-Min and Huyck, Charles},
  journal={Annals of Geophysics},
  volume={54},
  number={6},
  year={2012}
}

@article{plank2014rapid,
  title={Rapid damage assessment by means of multi-temporal SAR—A comprehensive review and outlook to Sentinel-1},
  author={Plank, Simon},
  journal={Remote Sensing},
  volume={6},
  number={6},
  pages={4870--4906},
  year={2014},
  publisher={Multidisciplinary Digital Publishing Institute}
}

@article{pesaresi2007rapid,
  title={Rapid damage assessment of built-up structures using VHR satellite data in tsunami-affected areas},
  author={Pesaresi, M and Gerhardinger, A and Haag, F},
  journal={International Journal of Remote Sensing},
  volume={28},
  number={13-14},
  pages={3013--3036},
  year={2007},
  publisher={Taylor \& Francis}
}

@inproceedings{alam2018crisismmd,
title = "CrisisMMD: Multimodal twitter datasets from natural disasters",
author = "Firoj Alam and Ferda Ofli and Muhammad Imran",
year = "2018",
month = "1",
day = "1",
language = "English",
pages = "465--473",
booktitle = "Proc. of the 12th ICWSM, 2018",
publisher = "AAAI press",
}

@article{ofli2016combining,
  title={Combining human computing and machine learning to make sense of big (aerial) data for disaster response},
  author={Ofli, Ferda and Meier, Patrick and Imran, Muhammad and Castillo, Carlos and Tuia, Devis and Rey, Nicolas and Briant, Julien and Millet, Pauline and Reinhard, Friedrich and Parkan, Matthew and others},
  journal={Big data},
  volume={4},
  number={1},
  pages={47--59},
  year={2016},
  %publisher={Mary Ann Liebert, Inc. 140 Huguenot Street, 3rd Floor New Rochelle, NY 10801 USA}
}

@inproceedings{imran2014aidr,
  title={{AIDR}: Artificial intelligence for disaster response},
  author={Imran, Muhammad and Castillo, Carlos and Lucas, Ji and Meier, Patrick and Vieweg, Sarah},
  booktitle={Proc. of the ACM Conference on WWW},
  pages={159--162},
  year={2014},
  %organization={ACM}
}

@article{imran2015processing,
  title={Processing social media messages in mass emergency: A survey},
  author={Imran, Muhammad and Castillo, Carlos and Diaz, Fernando and Vieweg, Sarah},
  journal={ACM Computing Surveys},
  volume={47},
  number={4},
  pages={67},
  year={2015},
  publisher={ACM}
}

@inproceedings{peterson2019when,
  title={When Official Systems Overload: A Framework for Finding Social Media Calls for Help during Evacuations},
  author={Peterson, Steve and Stephens, Keri and Hughes, Amanda and Purohit, Hemant},
  booktitle={Proceedings of the Information Systems for Crisis Response and Management Conference},
  pages={867--875},
  year={2019}
}

@article{hughes2009twitter,
  title={Twitter adoption and use in mass convergence and emergency events},
  author={Hughes, Amanda Lee and Palen, Leysia},
  journal={International Journal of Emergency Management},
  volume={6},
  number={3-4},
  pages={248--260},
  year={2009},
  publisher={Inderscience Publishers}
}

@book{castillo2016big,
  title={Big Crisis Data},
  author={Castillo, Carlos},
  year={2016},
  publisher={Cambridge University Press},
  location={Cambridge, UK}
}

@inproceedings{starbird2010chatter,
  title={Chatter on the red: what hazards threat reveals about the social life of microblogged information},
  author={Starbird, Kate and Palen, Leysia and Hughes, Amanda L and Vieweg, Sarah},
  booktitle={ACM Conference on Computer Supported Cooperative Work},
  pages={241--250},
  year={2010},
}

@inproceedings{imran2013extracting,
  title={Extracting information nuggets from disaster-related messages in social media},
  author={Imran, Muhammad and Elbassuoni, Shady Mamoon and Castillo, Carlos and Diaz, Fernando and Meier, Patrick},
  booktitle={Proc. of the 12th ISCRAM},
  year={2013}
}

@inproceedings{chen2013understanding,
  title={Understanding and classifying image tweets},
  author={Chen, Tao and Lu, Dongyuan and Kan, Min-Yen and Cui, Peng},
  booktitle={ACM International Conference on Multimedia},
  pages={781--784},
  year={2013},
  %organization={ACM}
}

@inproceedings{daly2016mining,
  title={Mining and Classifying Image Posts on Social Media to Analyse Fires},
  author={Daly, Shannon and Thom, J},
  booktitle={Proc. of the 13th ISCRAM},
  pages={1--14},
  year={2016},
  %organization={ISCRAM Association}
}

@article{simonyan2014very,
  title={Very deep convolutional networks for large-scale image recognition},
  author={Simonyan, Karen and Zisserman, Andrew},
  journal={arXiv preprint arXiv:1409.1556},
  year={2014}
}

@inproceedings{hiltz2013dealing,
  title={Dealing with information overload when using social media for emergency management: Emerging solutions.},
  author={Hiltz, Starr Roxanne and Plotnick, Linda},
  booktitle={Proc. of the 10th ISCRAM, 2013},
  year={2013},
  organization={ISCRAM}
}

@article{TurkerM:IJRS04,
author = {Turker, M. and San, B. T.},
title = {Detection of collapsed buildings caused by the 1999 Izmit, Turkey earthquake through digital analysis of post-event aerial photographs},
journal = {International Journal of Remote Sensing},
volume = {25},
number = {21},
pages = {4701-4714},
year = {2004},
doi = {10.1080/01431160410001709976},
%URL = {http://dx.doi.org/10.1080/01431160410001709976},
%eprint = {http://dx.doi.org/10.1080/01431160410001709976},
abstract = { In this study, the post-earthquake aerial photographs were digitally processed and analysed to detect collapsed buildings caused by the Izmit, Turkey earthquake of 17 August 1999. The selected area of study encloses part of the city of Golcuk, which is one of the urban areas most strongly hit by the earthquake. The analysis relies on the idea that if a building is collapsed, then it will not have corresponding shadows. The boundaries of the buildings were available and stored in a Geographical Information System (GIS) as vector polygons. The vector building polygons were used to match the shadow casting edges of the buildings with their corresponding shadows and to perform analyses in a building-specific manner. The shadow edges of the buildings were detected through a Prewitt edge detection algorithm. For each building, the agreement was then measured between the shadow producing edges of the building polygons and the thresholded edge image based on the percentage of shadow edge pixels. If the computed percentage value was below a preset threshold then the building being assessed was declared as collapsed. Of the 80 collapsed buildings, 74 were detected correctly, providing 92.50\% producer's accuracy. The overall accuracy was computed as 96.15\%. The results show that the detection of the collapsed buildings through digital analysis of post-earthquake aerial photographs based on shadow information is quite encouraging. It is also demonstrated that determining the optimum threshold value for separating the collapsed from uncollapsed buildings is important. },
}

@Article{FengT:NHESS14,
AUTHOR = {Feng, T. and Hong, Z. and Fu, Q. and Ma, S. and Jie, X. and Wu, H. and Jiang, C. and Tong, X.},
TITLE = {Application and prospect of a high-resolution remote sensing and geo-information system in estimating earthquake casualties},
JOURNAL = {Natural Hazards and Earth System Sciences},
VOLUME = {14},
YEAR = {2014},
NUMBER = {8},
PAGES = {2165--2178},
%URL = {http://www.nat-hazards-earth-syst-sci.net/14/2165/2014/},
DOI = {10.5194/nhess-14-2165-2014}
}

@article{fernandez2015uav,
	Author = {Fernandez Galarreta, J and Kerle, N and Gerke, M},
	Journal = {Natural Hazards and Earth System Sciences},
	Number = {6},
	Pages = {1087--1101},
	Publisher = {Copernicus GmbH},
	Title = {UAV-based urban structural damage assessment using object-based image analysis and semantic reasoning},
	Volume = {15},
	Year = {2015}}

@inproceedings{imran2014coordinating,
  title={Coordinating human and machine intelligence to classify microblog communications in crises.},
  author={Imran, Muhammad and Castillo, Carlos and Lucas, Jesse and Meier, Patrick and Rogstadius, Jakob},
  booktitle={ISCRAM},
  year={2014}
}

@inproceedings{nguyen2017automatic,
	Author = {Nguyen, Dat Tien and Alam, Firoj and Ofli, Ferda and Imran, Muhammad},
	Booktitle = {International Conference on Information Systems for Crisis Response and Management (ISCRAM)},
	Title = {Automatic Image Filtering on Social Networks Using Deep Learning and Perceptual Hashing During Crises},
	Month = {May},
	Year = {2017}}

@article{purohit2014identifying,
  title={Identifying seekers and suppliers in social media communities to support crisis coordination},
  author={Purohit, Hemant and Hampton, Andrew and Bhatt, Shreyansh and Shalin, Valerie L and Sheth, Amit P and Flach, John M},
  journal={Computer Supported Cooperative Work (CSCW)},
  volume={23},
  number={4-6},
  pages={513--545},
  year={2014},
  publisher={Springer}
}

@INPROCEEDINGS{Nattari:DSAA17, 
author={N. Attari and F. Ofli and M. Awad and J. Lucas and S. Chawla}, 
booktitle={2017 IEEE International Conference on Data Science and Advanced Analytics (DSAA)}, 
title={{Nazr-CNN: Fine-Grained Classification of UAV Imagery for Damage Assessment}}, 
year={2017}, 
volume={}, 
number={}, 
pages={50-59}, 
doi={10.1109/DSAA.2017.72}, 
ISSN={}, 
month={Oct},}
